\pdfoutput=1
\documentclass[a4paper,12pt,english]{article}

\usepackage[pdftex]{graphicx}
\usepackage{graphicx}
\usepackage{colordvi}
\usepackage{fancybox}
\usepackage{cancel}
\usepackage{amsmath}
\usepackage{amssymb}
\usepackage{caption}
\usepackage{appendix}
\usepackage{hyperref} %naredi dokument aktiven (v DVI in PDF)
\usepackage{multirow}
\hypersetup{colorlinks=true, linkcolor=black, urlcolor=blue}
\usepackage{float}
\usepackage{cite}

\usepackage{wasysym}

\usepackage{color}
\definecolor{gray}{rgb}{0.5,0.5,0.5}

\newcommand\lsim{\mathrel{\rlap{\lower4pt\hbox{\hskip1pt$\sim$}}
    \raise1pt\hbox{$<$}}}
\newcommand\gsim{\mathrel{\rlap{\lower4pt\hbox{\hskip1pt$\sim$}}
    \raise1pt\hbox{$>$}}}

\newcommand{\beq}{\begin{equation}}
\newcommand{\eeq}{\end{equation}}
\newcommand{\bea}{\begin{eqnarray}}
\newcommand{\eea}{\end{eqnarray}}
\newcommand{\bem}{\begin{pmatrix}}
\newcommand{\eem}{\end{pmatrix}}
\newcommand{\noi}{\noindent}

\begin{document}

\numberwithin{equation}{section}

\begin{flushright}
CP3-Origins-2016-046 DNRF90
%December, 2015
\end{flushright}

\bigskip

\begin{center}

{\Large\bf  
Asymptotically Safe Grand Unification}
\vspace{1cm}

\centerline{Borut Bajc$^{a,}$\footnote{borut.bajc@ijs.si} and 
Francesco Sannino$^{b,c,}$\footnote{sannino@cp3.dias.sdu.dk}}

\vspace{0.5cm}
\centerline{$^{a}$ {\it\small J.\ Stefan Institute, 1000 Ljubljana, Slovenia}}
\centerline{$^{b}$ {\it\small CP$^3$-Origins \& the Danish IAS, University of Southern Denmark, }}
\centerline{{\it\small Campusvej 55, DK-5230 Odense M, Denmark}}
\centerline{$^{c}$ {\it\small Universit\'e de Lyon, France, Universit\'e Lyon 1, CNRS/IN2P3, UMR5822 IPNL}}
\centerline{\it\small 
F-69622 Villeurbanne Cedex, France. }
\end{center}

\bigskip

\begin{abstract}
Phenomenologically appealing supersymmetric grand unified theories have large gauge 
representations and thus are not asymptotically free. Their ultraviolet validity is limited by 
the appearance of a Landau pole well before the Planck scale.  One could hope that these theories save themselves, before the inclusion of gravity, by generating an interacting ultraviolet fixed point, similar to the one recently discovered in non-supersymmetric gauge-Yukawa theories. Employing a-maximization, a-theorem, unitarity bounds, as well as positivity of other central charges we nonperturbatively rule out this possibility for a broad class of prime candidates of phenomenologically relevant supersymmetric grand unified theories. We also uncover candidates passing 
these tests, which have either exotic matter or contain one field decoupled from 
the superpotential.  The latter class of theories contains a model with 
the minimal matter content required by phenomenology. 
  \end{abstract}

\clearpage

\tableofcontents

\hypersetup{colorlinks=true, linkcolor=red, urlcolor=blue}

\section{Introduction}
Theories of grand unification continue to play an important role as guiding principle when searching for extensions of the Standard Model.  They offer a natural explanation for the observed  
quantization of the electric charge \cite{Pati:1974yy,Georgi:1974sy}, and predict the unification of the distinct SM gauge couplings at high 
energy \cite{Georgi:1974yf}. The latter occurs when adding to the SM specific matter transforming according to incomplete representations of the grand unified theory (GUT). 

Supersymmetry is a natural playground for the unification scenario since it almost automatically predicts the correct low energy spectrum that allows for one step-unification of the 3 gauge couplings. These meet at approximately 
$2\times10^{16}$ GeV \cite{Dimopoulos:1981yj,Ibanez:1981yh,Einhorn:1981sx,Marciano:1981un}. Moreover the different low energy matter fields, of a given generation, also unify in a larger representation of the gauge group, i.e. the 16 (27) of SO(10) \cite{so10} (or $E_6$ \cite{Gursey:1975ki}). This also, in turn, predicts new states such as the occurrence of a right-handed neutrino (plus extra vector-like matter in $E_6$) that fits naturally in the  see-saw mechanism \cite{seesaw}.

It is a fact, however, that asymptotic freedom is not always respected in supersymmetric GUTs such as the ones that predict exact R-parity conservation \cite{Mohapatra:1986su,Font:1989ai,Martin:1992mq} 
at low energy \cite{Aulakh:1997ba,Aulakh:1998nn,Aulakh:1999cd}. The reason being that one needs large matter representations \cite{Aulakh:1982sw,Clark:1982ai,Babu:1992ia,Aulakh:2003kg} under SO(10). This means that the coefficient of the one-loop gauge beta function

\beq
\beta_{1-loop}=3T(G)-\sum_iT(R_i)
\eeq

\noi
is strongly negative leading to a Landau pole typically just above the GUT scale but comfortably below the canonical gravity scale (for a 2-loop study see \cite{Aulakh:2015efa}).  Embedding SO(10) in larger gauge groups, for example 
$E_6$ cannot help \cite{Bajc:2013qra,Babu:2015psa} because the resulting theory is even less asymptotically free. 

One could envision different ways to go around this issue, for example one could push the unification scale closer to the gravity one via specific threshold corrections and hope that gravity will work its magic. 

Another appealing possibility is that these theories save themselves, before gravity sets in, by developing an ultraviolet interacting fixed point in all couplings. 

The hope for such a possibility stems from the discovery  \cite{Litim:2014uca}  that vector-like non asymptotically free gauge-Yukawa theories can indeed be fundamental theories at all scales\footnote{An important aspect of asymptotic safety in perturbative gauge-Yukawa theories is that scalars are required to tame the gauge fluctuations \cite{Litim:2014uca,Litim:2015iea,Bond:2016dvk}. Earlier investigations  of perturbative IR and UV interacting fixed points for gauge-Yukawa theories  \cite{Antipin:2013pya} were instrumental for the discovery in \cite{Litim:2014uca}.  Asymptotic safety might occur also without elementary scalars \cite{Pica:2010xq,Holdom:2010qs} but it would require a phase transition in the number of matter fields \cite{SanninoERG2016}. }. The situation  changes  when considering the supersymmetric cousins of the theory investigated in  \cite{Litim:2014uca}. It was, in fact, demonstrated in \cite{Intriligator:2015xxa} that these supersymmetric cousins are unsafe, along with a much broader class of supersymmetric theories, further extending the one in \cite{Martin:2000cr}. The first study of asymptotically safe chiral gauge theories, some of which resembling GUT-like non-supersymmetric theories, appeared  in \cite{Molgaard:2016bqf} while semi-simple gauge groups in \cite{Esbensen:2015cjw}. Asymptotic safety has been invoked by Weinberg \cite{Weinberg:1980gg} to tame quantum gravity  \cite{Niedermaier:2006ns,Niedermaier:2006wt,
Percacci:2007sz,Reuter:2012id,Litim:2011cp} \footnote {UV conformal extensions of the standard model with and without gravity have also been discussed in the literature 
%\cite{Kazakov:2002jd} -\cite{Meibohm:2015twa}}
\cite{Kazakov:2002jd,Gies:2003dp,Morris:2004mg,
Fischer:2006fz,
 Fischer:2006at,
Kazakov:2007su,Zanusso:2009bs,
Gies:2009sv,Daum:2009dn,Vacca:2010mj,Folkerts:2011jz,
Bazzocchi:2011vr,Gies:2013pma,Antipin:2013exa,Dona:2013qba,Bonanno:2001xi,Meissner:2006zh,Foot:2007iy,Hewett:2007st,Litim:2007iu,Shaposhnikov:2008xi,
Shaposhnikov:2008xb,Shaposhnikov:2009pv,Weinberg:2009wa,
Hooft:2010ac,
Hindmarsh:2011hx,Hur:2011sv,Dobrich:2012nv,Tavares:2013dga,
Tamarit:2013vda,Abel:2013mya,Antipin:2013bya,Heikinheimo:2013fta,
Gabrielli:2013hma,Holthausen:2013ota,Dorsch:2014qpa,Eichhorn:2014qka,Sannino:2014lxa,Nielsen:2015una,Codello:2016muj,Meibohm:2015twa}.}. 
 
 It is therefore timely and relevant to investigate the ultraviolet fate of a broad class of supersymmetric GUTs in which asymptotic freedom is lost. 

We start with a pedagogical introduction and description of the tools that we will use to uncover the dynamics of these theories. In particular we will investigate non asymptotically free SO(10) theories with different matter representations and with(out) superpotentials. Although we will show that a wide class of theories cannot abide all the constraints simultaneously we do find exotic theories featuring extremely large numbers of matter fields passing the tests. Besides the exotic models we also uncover a minimal model, with just 3 copies of $16$'s, as well as one representative for each 
of the $10$, $210$, $126$ and $\overline{126}$ multiplets of SO(10) that can still be asymptotically safe. 

We structure our paper as follows: In Section \ref{tests} we briefly review, for the benefit of the reader,  the rationale behind the full set of field-theoretical constraints we will use to discriminate among the possible candidate fixed points we will analyse. Section \ref{SO(10)} opens with a brief self-contained introduction and justification of SO(10) grand unified models featuring several matter representations. The rest of the section is devoted to the analysis of a broad class of SO(10) grand unified models with and without superpotentials. We finally offer our conclusions in section \ref{conclusions}. 

\section{Consistency Checks and Constraints } 
\label{tests}
 
Since the grand unified theories investigated here supersymmetric we have  a number of consistency checks and general constraints at our disposal to analyse the potential existence of any RG fixed point.   If such a RG fixed points exists in an ${\cal N} =1$ superconformal field theory (SCFT) it will necessarily possess a conserved $U(1)_R$ global symmetry.  Furthermore the $U(1)_R$ current is in the same supermultiplet \cite{Ferrara:1974pz} as the energy-momentum tensor and the supercharge currents; this leads to several exact relations and constraints that we briefly review in this section.

\subsection{Unitary constraints}
For a unitary theory, the operators form unitary representations of the superconformal group, which implies that operator dimensions have various lower bounds.  For example, regardless of supersymmetry, all gauge invariant spin $j=\bar j=0$ operators have the lower bound (generators act with implicit commutators) \cite{Mack:1975je} (see also e.g. \cite{Grinstein:2008qk})
\beq\label{dimone}
D({\cal O})\geq 1, \qquad D({\cal O})=1 \leftrightarrow P_\mu P^\mu ({\cal O})=0,
\eeq
so the bound is saturated if and only if the operator ${\cal O}$ is a decoupled, free field.  Chiral primary operators have dimension, $D$, and superconformal $U(1)_R$ charge, $R$, related by 
\beq\label{chiralprimary}
D({\cal O})=\frac{3}{2} R({\cal O}).
\eeq
Using \eqref{chiralprimary} for the matter chiral superfields $Q_i$ one can relate the matter anomalous dimensions $\gamma _i$ to their superconformal $U(1)_R$ charge.  
\begin{equation}
D(Q_i)\equiv 1+\frac{1}{2}\gamma _i(g)=\frac{3}{2}R(Q_i) \equiv \frac{3}{2} R_i .\label{aba:eq5}
\end{equation}

\subsection{Central charges and their positivity}

We summarise here the constraints due to the positivity of the coefficients related to the stress-energy trace anomaly. These have been derived by considering the effects of an external supergravity background for theories with sources for conserved flavor currents  stemming from trace anomaly and proportional to the square of the dual of the Riemann curvature, the square of the Weyl tensor, as well as the square of the flavor symmetry field strength. These functions of the $R$ charge are indicated respectively with $a(R)$, $c(R)$ and $b(R)$  \cite{Anselmi:1997am,Anselmi:1997ys}.

The conformal anomaly $a$ of the SCFT is exactly given by the superconformal $U(1)_R$ 't Hooft anomalies  \cite{Anselmi:1997am,Anselmi:1997ys} (we rescale the overall normalization factor of 3/32 for convenience)
\beq\label{aanselmi}
a(R)=3{\rm Tr}U(1)_R^3-{\rm Tr}U(1)_R.
\eeq
Let's determine this function for a gauge theory with gauge group, $G$, and matter fields $Q_i$, in representations $r_i$ of $G$, the 't Hooft anomalies evaluate to 
\beq\label{agaugetheoryderivation}
a(R) = |G| \left[ 3 R_V^3 - R_V\right] + \sum_i   | r_i| \left[ 3 (R_i-1)^3  - (R_i - 1) \right]  = 2|G|+\sum _i |r_i| a_1(R_i)
\eeq
where $ |G| = r_{\rm adjoint}$ is the number of generators in the adjoint representation and   $| r_{i}|$  is the dimension of the representation $r_{i}$.  We must use in $a(R)$   the fermion $R$ charges that for the gluino is exactly $R_V = R(V) = 1$ with $V$ the vector chiral superfield, while for each chiral superfield $Q = \phi_{Q} + \sqrt{2} q\theta + \theta^2 F_{Q}$ we have $R(q) = R(Q) - 1$ because $R(\theta)=-1$, and we define the function 
\begin{equation}
a_1(R)\equiv 3(R-1)^3-(R-1) = (1-R) \left[1 - 3(1-R)^2 \right] ~. \label{aRex}
\end{equation}
\bigskip 

The c-function reads  \cite{Anselmi:1997am,Anselmi:1997ys} 
\beq\label{canselmi1}
c(R)=9{\rm Tr}U(1)_R^3-5{\rm Tr}U(1)_R.
\eeq

For a generic gauge theory we have: 
\beq\label{canselmi2}
c(R)= |G|(9 - 5) + \sum_i |r_i| \left[9(R_i -1)^3 - 5(R_i -1)\right] = 4|G| + \sum_i |r_i| (1 - R_i) \left[5 - 9(1-R_i)^2\right] \ ,
\eeq
and we dropped the overall normalization factor of 1/32. 

\bigskip 

The flavor b-function reads  \cite{Anselmi:1997am,Anselmi:1997ys} 
\beq\label{banselmi}
b(R)={\rm Tr}U(1)_R F^2 =  \sum_i |r_i| (1 - R_i)F^2_i \ .
\eeq
We have dropped the overall normalization factor of 3 and $F_i$ are the flavor charges for each representation.

\subsection{$a$-maximization}

\begin{figure}[ht]
\begin{center}
\includegraphics[width=.5\textwidth]{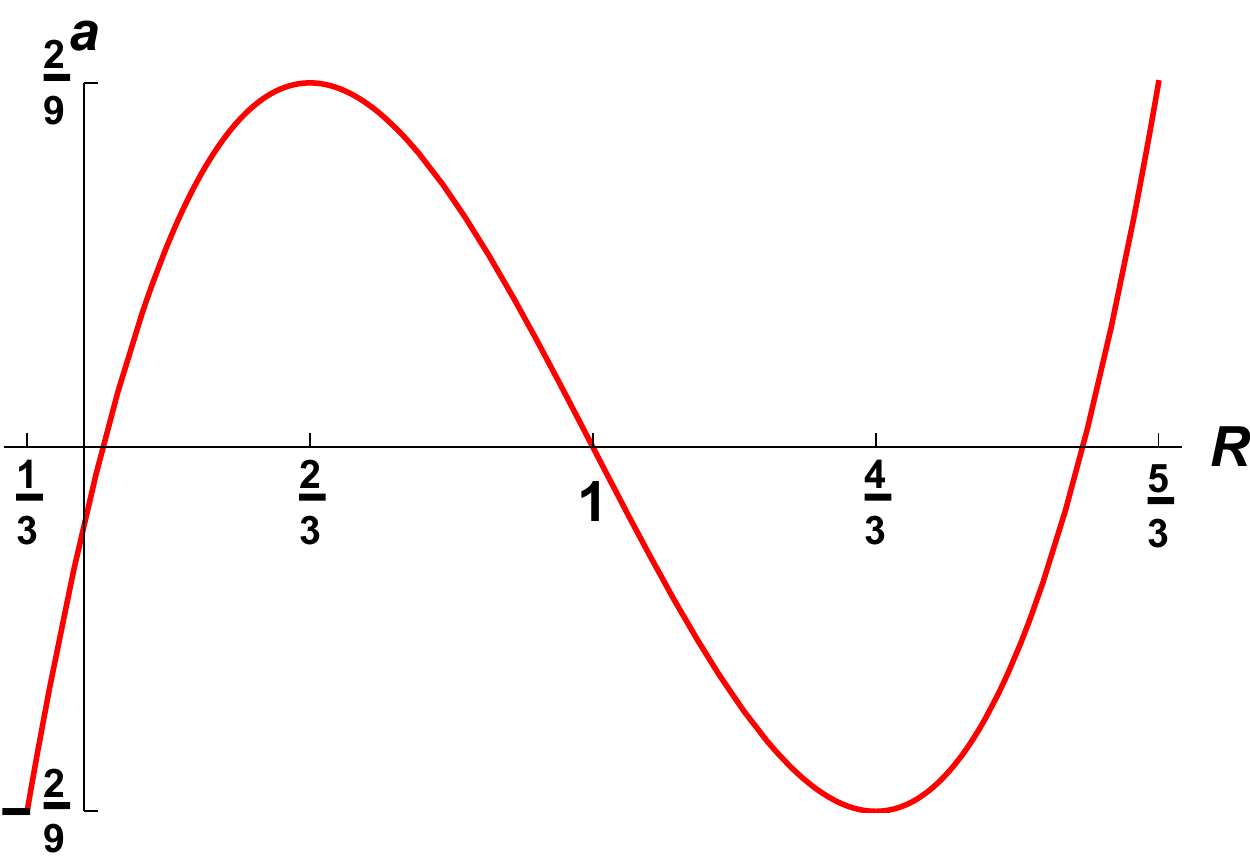} 
  \caption{The function  $a_{1}(R)$. }
\bigskip\noindent
\label{agraph}
\end{center}
\end{figure}

Among all possible, conserved $U(1)_R$ symmetries, the superconformal $U(1)_R$ is the one maximizing  $a(R)$ pioneered in \cite{Intriligator:2003jj}.  For example, for a chiral superfield $X$ of charge $R(X)=R$ (so $R(\psi _X)=R-1$), the function is $a(R)=a_1(R)$ in \eqref{aRex}.  The function $a_1(R)$ has a local maximum at the free-field value, $R={2\over 3}$, and a local minimum at $R={4\over 3}$, see Fig.~\ref{agraph}.  % Indeed $a_1(R)=-a_1(2-R)$, so $a_1(R=1)=0$, that it is in line with massive operators contributing $a=0$.   The latter statement comes from the fact that a mass term operator is quadratic in the potential and therefore the field must carry unit R charge.    
In addition $a_1(R)$ is below the local maximum, $a_1(R)<a_1(R=2/3)$ for $R<5/3$. (see \cite{Intriligator:2005if} for a further related phase diagnostic).
 We  maximize the function \eqref{aRex}  for unconstrained, i.e. free chiral superfields and obtain  $R_*=2/3$, which is the free-field value of the R-charge, corresponding to $D (X)=1$.  When interactions are present, we maximize $a(R)$ requiring the interactions to preserve the R-symmetry. Accidental symmetries, if present, affect a-maximization \cite{Kutasov:2003iy,Intriligator:2003mi} yielding a larger value of $a$. 

\subsection{Beta functions}
 Beta functions are proportional to how the couplings break the superconformal $U(1)_R$ when going away from the fixed point.  The supersymmetric gauge coupling beta function embodies a remarkable property, it is proportional to the ABJ triangle anomaly of the $U(1)_R$ current with two $G$ gauge fields, i.e. ${\rm Tr}~G^2U(1)_R$:
\beq\label{gaugebeta}
\beta (g)=-\frac{3g^3}{16\pi ^2}f(g^2){\rm Tr}~ G^2U(1)_R, \qquad 
{\rm Tr}~G^2U(1)_R=T(G)+\sum _i T(r_i)(R_i-1).
\eeq
Our normalization for the quadratic Casimir of the adjoint $T(G)$ is  $T(SU(N_c))=N_c$, so that the fundamental representation of $SU(N)$ has $T(r_{\rm fund})=\frac{1}{2}$. The function $f(g^2)=1+{\cal O}(g^2)$ is scheme dependent (and presumed positive).   The above \eqref{aba:eq5} is the NSVZ exact beta function \cite{Novikov:1983uc}, in which a specific scheme is employed for $f(g^2)$:
\beq\label{gaugebetaNSVZ}
\beta (8\pi ^2g^{-2})=f(g^2)(3T(G)-\sum _i T(r_i)(1-\gamma _i(g))).
\eeq
For superpotential terms with trilinear interactions $W_y$, the beta function for the holomorphic coupling $y$ reads
\beq\label{betah}
\beta (y)=\frac{3}{2} y (R(W_y)-2).
\eeq

\subsection{$a$, $b$  and $c$-theorems}

For any super CFT not only these coefficients must be positive \cite{Anselmi:1997am,Anselmi:1997ys} but it is also expected, following Cardy's conjecture, a 4d version of the $a$-theorem \cite{Cardy:1988cwa,Komargodski:2011vj,Komargodski:2011xv,Zamolodchikov:1986gt,Osborn:1989td,Jack:1990eb}, that reads 
\beq\label{athm}
\Delta a \equiv a_{UV}-a_{IR}>0.
\eeq
For free theories these coefficients are automatically positive. This implies that for asymptotically free theories they are automatically positive at the trivial UV fixed point while for asymptotically safe theories they are automatically positive in the infrared. In fact, the free value for gauge theories reads: 
\beq
\label{freecharges}
a_{\rm free} = 2 |G| + \frac{2}{9}\sum_i |r_i| \ , \quad c_{\rm free} = 4 |G| +  \frac{4}{3}\sum_i |r_i| \ , \quad b_{\rm free} = \frac{1}{3}\sum_i |r_i| F^2_i 
\eeq 
where $R_i = 2/3$ because all the chiral superfields are free, i.e.  $D_i = 1$, as it should be for a non-interacting field. It is worth mentioning that the physical dimension of the vector chiral superfield is always (also in the interacting theory) the free one since the $R$ charge of the gluino is fixed and $D(V) =  3/2 R(V) = 3/2 = D(\lambda)$ with $\lambda$ the gluino.  For the free theory we do not care about anomaly free value for $R$ charges because there are no interactions.

Interesting constraints emerge when requiring positivity of $b_{IR(UV)}$ and  $c_{IR(UV)}$ for the interacting IR(UV) fixed point in asymptotically free (safe)  field theories along with the $\Delta a >0$ condition. At the interacting fixed point the only $R$ charges that matter are the ones that allow for the interacting field theory to be consistent, including 't Hooft anomaly (free) conditions and superpotential constraints. The above implies: 
\beq
\label{interacting-positivity}
c_{\rm FP} =  4|G| + \sum_i |r_i| (1 - R_i) \left[5 - 9(1-R_i)^2\right]  >0 \ , \quad  b_{\rm FP} =  \sum_i |r_i| (1 - R_i)F^2_i >0\ ,
\eeq
and 
\beq
\label{a-th}
\Delta a   =  a_{\rm UV} - a_{\rm IR} =  \pm \left(  2|G|+\sum _i |r_i| a_1(R_i) - a_{\rm free}\right) = \pm\frac{1}{9}\sum _i |r_i|\left[ (3R_i - 2)^2 (3R_i - 5)\right] >0\ ,
\eeq
where the plus(minus) sign corresponds to the asymptotically safe(free) interacting fixed point. The constraint in \eqref{a-th} is stronger for asymptotically safe theories \cite{Martin:2000cr,Intriligator:2015xxa} since it requires at least one chiral superfield to have a quite sizable $R$ charge larger than  $5/3$. Assuming, for example, the presence of an asymptotically safe fixed point for super QCD once asymptotic freedom is lost, i.e. $N_f>3N_c$, one discovers that $R_Q = R_{\widetilde{Q} }= 1-N_c/N_f$ assume values between $1$ and $2/3$ and therefore the theory violates the constraint in \eqref{a-th}  \cite{Martin:2000cr,Intriligator:2015xxa} while it still respects positivity of the remaining constraints. An IR interacting fixed point, relevant for super QCD conformal window \cite{Seiberg:1994pq,Intriligator:1995au},  in asymptotically free field theories, on the other hand, requires the milder condition $R_i< 5/3$. 

\subsection{Tracking the $R$-charge without the superpotential}

With vanishing superpotential, the $a$-function is defined as \cite{Kutasov:2003ux,Kutasov:2004xu}

\beq
a(R_i,\lambda_G)=2|G|+\sum_i|r_i|\left(3(R_i-1)^3-(R_i-1)\right)+
\lambda_G\left(T(G)+\sum_iT(r_i)(R_i-1)\right)
\eeq

\noi
where $\lambda_G$ is the Lagrange multiplier which enforces the vanishing of the 
NSVZ $\beta$-function at the superconformal fixed point. From

\beq
\frac{\partial a(R_i,\lambda_G)}{\partial R_i}=0
\eeq

\noi
one finds 

\beq
\label{Ri}
R_i=1-\frac{\epsilon_i}{3}\sqrt{1-\frac{T(r_i)\lambda_G}{|r_i|}}
\eeq

\noi
with $\epsilon_i^2=1$. Reality of $R_i$ requires

\beq
\label{lambdaGmin}
\lambda_G\leq \lambda_G^{max}\equiv min_i\left(\frac{|r_i|}{T(r_i)}\right)
\eeq

\vskip 1cm  
We will also presume that the interacting fixed point is smoothly connected to the non-interacting fixed point when the coupling vanishes. This allows to enforce continuity of the $R$ charges. 

We are now ready to investigate the dynamics of grand unified theories that are not asymptotically free. 

\section{\label{SO(10)} Can SO(10) GUT  be  asymptotically safe?}

We will now use the above machinery to investigate whether SO(10) GUT theories can be asymptotically safe rather than free. We first summarize how and why the loss of asymptotic freedom appears when trying to construct  models that automatically embody $R$-parity. We then analyze whether these theories can nonperturbatively flow to an UV fixed point by applying the above tests.  

\subsection{Gaining $R$ parity by loosing asymptotic freedom}
We mentioned in the introduction that non asymptotically free grand unifications can provide a rationale for the existence of low energy $R$ parity \cite{Clark:1982ai,Babu:1992ia,Aulakh:2003kg}.  The latter stems from the SO(10) Cartan subalgebra 
generator $B-L$ through
\beq
R=(-1)^{3(B-L)+2S}=M(-1)^{2S}  \quad {\rm with} \quad  M=(-1)^{3(B-L)} \ .
\eeq
We see that $R$ parity, up to the spin $S$, identifies with the matter parity $M$. An elegant way to break the rank of SO(10) without breaking spontaneously the $R$-parity is to introduce a Higgs sector transforming according to the $126+\overline{126}$  dimensional representation\footnote{We need simultaneously $126$ and $\overline{126}$  to cancel 
the D-terms.} of SO(10).  Indeed in the ${126}$ ($\overline{126}$) the only possible SM and SU(5) singlet has $B-L=-2(2)$ that preserves $R$-parity.   

Since in SO(10)

\beq
16\times16=10+126+120 \ ,
\eeq

\noi
the Yukawa couplings (and thus all SM fermion masses) could arise via the following linear combination:
\beq
W_{Yukawa}=16_a\left(Y_{10}^{ab}10+Y_{126}^{ab}\overline{126}+Y_{120}^{ab}120\right)16_b \ , 
\eeq
\noi
with $a,b=1,2,3$ running over the generations. From SO(10) one can show that 

\beq
Y_{10,126}=+Y_{10,126}^T\;\;\;,\;\;\;Y_{120}=-Y_{120}^T \ .
\eeq
In fact a minimal choice to generate  realistic mixings among the generations, i.e. the physical $V_{CKM}$ and $V_{PMNS}$ matrices, is to add to  the already present $\overline{126}$ the $10$ dimensional representation alone \footnote{The other choice, i.e.  $\overline{126}+120$, does not have enough parameters to reproduce the physical results. The reason being the antisymmetric nature of  the matrix $Y_{120}$.}.   

For successful model building two requirements must still be met: First, we need to break SO(10) down to the SM gauge group which cannot be accomplished by the $\overline{126}$ that at most can break it to SU(5); Second, the MSSM Higgses must be contained in both the $10$ and $\overline{126}$. Both problems can be addressed by  introducing the $210$ representation. The reasons being that: The 3 SM singlets of $210$ are enough to further break SU(5) to the SM group; The renormalizable operator $210\,10\,\overline{126}$ sources via a nonzero doublet vev in $10$  a related vev in $\overline{126}$. 

To summarize, the minimal SO(10) model we will consider is composed of
\beq
3\times16+126+\overline{126}+10+210 \ ,
\eeq
that in the end yields a non-asymptotically free theory with the following extremely large coefficient of the one-loop beta function
\beq
\beta_{1-loop}=-109 \ ,
\eeq
implying that a Landau pole is reached very quickly and below the Planck scale.  The  emergence of an interacting ultraviolet fixed point could save the theory. We will therefore investigate such a possibility in the next session. 

\subsection{SO(10) GUT without superpotential is unsafe}

   We commence our analysis by demonstrating that: {\it Minimal SO(10) with $ 3\times16+126+\overline{126}+10+210 $ matter content and  vanishing 
superpotential does not have a UV fixed point.}

To prove this we start with the NSVZ beta function 
\beq
\beta_{NSVZ}(\lambda_G)\equiv T(G)+\sum_iT(r_i)(R_i(\lambda_G)-1) \ ,
\eeq

\noi
in which $R_i(\lambda_G)$ is given by (\ref{Ri}).  We, of course, reproduce in the non-interacting IR limit ($\lambda_G=0$ and $\epsilon_i=+1$) 

\beq
\label{beta0}
\beta_{NSVZ}(0)=\beta_{1-loop}/3=-109/3 \ .
\eeq
 We use the Dynkin indices from \cite{Slansky:1981yr,Feger:2012bs}  summarized, for reader's convenience, for
the lowest dimensional representations in SO(10) in Table \ref{Dynkin}.
\begin{table}[H]
\centering
\begin{tabular}{ |c||c|c|c|c|c|c|c|c| } 
\hline
$\phi_i$ & 10 & 16 & 45 & 54 & 120 & 126 & 144 & 210 \\
\hline
$T(r_i)$ & 1 & 2 & 8 & 12 & 28 & 35 & 34 & 56 \\
\hline
 \end{tabular}
\caption{\label{Dynkin} Dynkin indices $T(r_i)$ for some low-dimensional representations $\phi_i$ 
of SO(10). 
}
\end{table}

Now, let's assume that an UV fixed point occurs nonperturbatively in the theory.  Because we require $a_{UV}>a_{IR}$  at least one $\epsilon_i=-1$ must be negative implying that by continuity in $\lambda_G$ we need to reach the point where
\beq
\lambda_G^{max}=min_i\left(\frac{|r_i|}{T(r_i)}\right)=\frac{|126|}{T(126)}=
\frac{|\overline{126}|}{T(\overline{126})}=\frac{126}{35} \ .
\eeq

However for this value of $\lambda_G$ we find 

\beq
\beta_{NSVZ}(\lambda_G^{max})=4-\sqrt{\frac{11}{5}}>0 \ ,
\eeq

\noi
and therefore the $\beta_{NSVZ}(\lambda_G)$ must have changed sign between $\lambda_G=0$ in the infrared and 
$\lambda_G=\lambda_G^{max}$. Assuming continuity an apparent fixed point exists 
$\lambda_G^* < \lambda_G^{max}$ for which\footnote{Here and in the following we round all real 
numbers to 3 digits.} 
\beq
\beta_{NSVZ}(\lambda_G^*)=0 \ , \quad {\rm and} \quad \lambda_G^* = 3.57 \ .
\eeq
  However for this value we find  
  \beq
a_{UV}=a(R_i(\lambda_G^{*})) =125<206 =a(R_i(0))=a_{IR} \ ,
\eeq
\noi
showing that the alleged fixed point violates the a-theorem constraint expressed in \eqref{a-th} and therefore cannot be physical. 
\vskip .5cm

We now move to consider a more general matter field content of SO(10) without superpotential and test whether one can achieve an acceptable UV fixed point. In practice we require:  

\begin{itemize}

\item[1)]{No zero to appear in the $\beta_{NSVZ}(\lambda_G)$ for the branch connected to the 
perturbative IR region (i.e. with all $\epsilon_i=+1$) with $\lambda_G\leq\lambda_G^{max}$. 
Notice that $\lambda_G^{max}$ differs for different theories;}

\item[2)]{A possible UV zero in the $\beta_{NSVZ}(\lambda_G)$ to occur for, at least, some negative $\epsilon$'s, i.e. 
$\epsilon_k=-1$ with $\lambda_G\leq\lambda_G^{max}\equiv min_i(|r_i|/T(r_i))=|r_k|/T(r_k)$;}

\item  [3)]{This solution must either 
satisfy $a_{UV}>a_{IR}$, or develop at least one non-interacting gauge invariant operator (GIO) and thus by eliminating the operator has a chance for a  
modified $a$.}

\end{itemize}

We now perform a scan over the following two families of theories

\begin{itemize}

\item [i)]{We first consider the same type of matter fields $10$, $16$ and/or $\overline{16}$, $126$ and/or $\overline{126}$ and $210$ but scan over the theories featuring from $0$ up to $3$ copies of each field. The total number of cases is therefore 
$4^4-1=255$. Only 240 of these combinations have a negative 1-loop $\beta$-function and are thus interesting to investigate. Clearly our first example, the one discussed in detail at the beginning of this section, corresponds to a special case of this family of theories, i.e. 1 multiplet of $10$, 3 multiplets 
$16$, 2 multiplets $126$ (or $\overline{126}$) and one multiplet $210$. Among these 240 theories we found that 37 of them satisfy point 1) and 2) above.}

\item [ii)]{
In the second example we consider the fields $10$, $16$ (or $\overline{16}$), $45$, $54$, $120$ 
$126$ (or $\overline{126}$), $144$ (or $\overline{144}$), and $210$. We scan over all 
possibilities of having or not having each of these fields once. This means we consider 
$2^8-1=255$ different theories, of which again 240 have 
a negative 1-loop $\beta$-function. Of these theories 23 are found to satisfy point 1) and 2) above. }
 
\end{itemize}

We were unable to find acceptable asymptotically safe solutions for any of the $2\times240=480$ different models above satisfying simultaneously the conditions 1), 2) and 3).  These findings extend the results of  \cite{Intriligator:2015xxa}. Nevertheless exotic theories exist passing these tests such as the theory with 
$274909$ generations of $10$, and $5161$ generations of $126$ (part or all of them can be 
$\overline{126}$). The would be UV fixed point seems to occur for 
\bea
\lambda_G^*&=&-28.5 \ ,
\eea 
for which 
\bea 
a_{UV}-a_{IR}&=&1.17\times 10^4 \ ,\\
(R_{10},R_{126})&=&(0.346,2.00) \ ,\\
b_{UV}&=&2.99\times10^5 \times F_{10}^2\ ,\\
c_{UV}&=&4.60\times10^6 \ .
\eea
All constraints are met and no GIO becomes non-interacting.  There are other exotic solutions of this type,  with some containing 3 generations of the  $16$ matter. These solutions are far from phenomenologically viable while help elucidating the difficulty in constructing asymptotically safe supersymmetric quantum field theories.

\subsection{\label{W}Minimal model with a superpotential }

We now extend the analysis above to the case of a non-vanishing superpotential. We shall use here as well the continuity of the R-charges $R_i$ as functions of 
the Lagrange multiplier $\lambda_G$, and further add Lagrange multipliers $\lambda_a$ stemming from each new interaction in the superpotential $W$.

 Let's therefore consider all the permitted trilinear terms in the superpotential \cite{Aulakh:1982sw,Clark:1982ai,Babu:1992ia,Aulakh:2003kg}:
\beq
W=y_1\,210^3+y_2\,210\,126\,\overline{126}+y_3\,210\,126\,10+y_4\,210\,\overline{126}\,10+
\sum_{a,b=1,2,3}16_a\,16_b\,\left(y_{5,ab}\,10+y_{6,ab}\,\overline{126}\right)
\label{Wgeneral}
\eeq
The function $a$ assumes the form 
\bea
a&=&2|G|+\sum_i|r_i|a_1(R_i)+\lambda_G\left(T(G)+\sum_iT(r_i)\left(R_i-1\right)\right) \nonumber \\
&+&\lambda_1\left(2-3R_{210}\right)+\lambda_2\left(2-R_{210}-R_{126}-R_{\overline{126}}\right)
+\lambda_3\left(2-R_{210}-R_{126}-R_{10}\right) \nonumber\\
&+&\lambda_4\left(2-R_{210}-R_{\overline{126}}-R_{10}\right)+\sum_{a,b=1,2,3}\lambda_{5,ab}\left(2-R_{10}-R_{16_a}-R_{16_b}\right) \nonumber\\
&+&\sum_{a,b=1,2,3}\lambda_{6,ab}\left(2-R_{\overline{126}}-R_{16_a}-R_{16_b}\right) \ .
\label{ageneral}
\eea
%{\color{blue} We are now in the position to mimic the analysis performed earlier by requiring that at least one $R_i$ is larger than unity}. We summarise the cases where this cannot occur in Table \ref{forbiddenepsilon} \footnote{ We note that 
%  if some GIO become free at the fixed point, one should redo the analysis by  
%modifying the $a$-function as discussed later.}

%\begin{table}
%\centering
%\begin{tabular}{ |c|c| } 
%\hline
%terms in $W$ & fields which cannot have $\epsilon=-1$ \\
%\hline\hline
%- & $210,10,16_1,16_2,16_3$ \\
%6 & $210,10,16_1,16_3$ \\
%5 & $210,10,16_1,16_2$ \\
%5,6 & $210,10,16_1$ \\
%4 & $210,16_1,16_2,16_3$ \\
%4,6 & $210,16_1,16_3$ \\
%4,5 & $210,16_1,16_2$ \\
%4,5,6 & $16_1$ \\
%3 & $210,16_1,16_2,16_3$ \\
%3,5 & $210,16_1,16_2$ \\
%3,4 & $210$ \\
%2 & $10$ \\
%2,6 & $10$ \\
%2,5 & $10$ \\
%2,5,6 & $10$ \\
%1 & $10,16_1,16_2,16_3$ \\
%1,6 & $10,16_1,16_3$ \\
%1,5 & $10,16_1,16_2$ \\
%1,5,6 & $10,16_1$ \\
%1,4 & $16_1,16_2,16_3$ \\
%1,4,6 & $16_1,16_3$ \\
%1,4,5 & $16_1,16_2$ \\
%1,4,5,6 & $16_1$ \\
%1,3 & $16_1,16_2,16_3$ \\
%1,3,5 & $16_1,16_2$ \\
%1,2 & $10$ \\
%1,2,6 & $10$ \\
%1,2,5 & $10$ \\
%1,2,5,6 & $10$ \\
%\hline
% \end{tabular}
%\caption{ \label{forbiddenepsilon} Fields which cannot have $R>1$ by continuity argument in 
%theories defined by the superpotential terms in the first column.}
%\end{table}
%
%
%\subsection{The minimal model and its variations}

If all the trilinear terms are present then the matter $R$-charges are constrained to be the free ones, i.e. $R_i = 2/3$ for any $i$ forbidding a zero in the NSVZ beta function. A minimal approach is to remove just one field from the superpotential. It turns out that the best choice is one of the $16$ fields which we choose to be $16_1$ and the sum over $a,b$ in \eqref{Wgeneral} and \eqref{ageneral} go over 2 and 3.

  By extremizing the a-function we now obtain $R_{16_1}=113/6$ and all the others fields still possess $R=2/3$. The positivity 
requirements are satisfied, since

\beq
a_{UV}-a_{IR} = 2.72\times10^5> 0 \ ,\quad {\rm and} \quad c_{UV}=8.16\times10^5 > 0 \ .
\eeq

\noi
while there are no extra flavor symmetries. 

Therefore the present solution passes all known constraints needed by a superconformal fixed point.  By construction our solution describes a world with a decoupled massless generation. 
%There are two drawback though:
%
%\begin{itemize}
%
%\item
%The theory is invariant under a U(1) global phase rotation of $\psi_1$. Instantons 
%can generate gauge invariant superpotential terms of the form
%
%\beq
%W_{instantons}=\sum_{a>0}\sum_{\sum_ix_i=3+105a}c^{(a)}_{x}\Lambda^{-109a}
%16_1^{4a}\prod_{i\ne16_1}\phi_i^{x_i}
%\eeq
%
%The presence of any such operator would spoil the result we obtained and 
%thus the UV fixed point. We know of no reason why all these operators should vanish.
%
%!TEX encoding = UTF-8 Unicode\item
 %\end{itemize}

Our solution corresponds to a manifold of UV fixed points. In fact the $11$ equations 
(we consider here the sums over $a,b=2,3$) $\partial a/\partial\lambda=0$, which extremize the $a$-function 
(\ref{ageneral}), are expressed with only $7$ combinations $R_i(\lambda)$ (got previously from 
$\partial a/\partial R=0$). This means that from the numerical 
values of $R_i$ in the UV limit we can determine

\bea
\lambda_G&=&-2.29\times10^4\\
\lambda_1&=&-1.64\times10^5\\
\lambda_4&=&1.14\times10^4
\eea

\noi
and the following linear combinations

\bea
\lambda_2+\lambda_3&=&-8.01\times10^5\\
(\lambda_{5,22}+\lambda_{5,23}+\lambda_{5,33})-\lambda_2&=&7.67\times10^5\\
(\lambda_{6,22}+\lambda_{6,23}+\lambda_{6,33})+\lambda_2&=&-8.13\times10^5\\
(\lambda_{5,33}+\lambda_{6,33})-(\lambda_{5,22}+\lambda_{6,22})&=&0
\eea

The solution we found is thus not a fixed point, but a manifold of fixed points, 
somehow reminiscent (although with the role of UV and IR inverted) of the cases 
considered in \cite{Leigh:1995ep}.

\subsection{Gauge invariant fields becoming free}

If a singlet scalar gauge invariant operator of the chiral ring\beq
{\cal O}_\alpha=\prod_i\phi_i^{q_{\alpha i}} \ ,
\eeq
at the fixed point 
 \noi
acquires 
\beq
R_\alpha=\sum_iq_{\alpha i}R_i<2/3 \ ,
\eeq
unitarity is violated unless it becomes free. If this occurs one needs  to modify the $a$ function  
 accordingly \cite{Kutasov:2003iy}
\beq
a(R_i)\to a(R_i)+\sum_\alpha \left(a_1\left(2/3\right)-a_1\left(R_\alpha\right)\right) \ .
\eeq
The maximization of the modified $a$ function is subject to the same constraints discussed earlier. Note that these additional terms to the $a$-function naively tend to increase its value. 
This means that a candidate fixed point with $a_{UV}<a_{IR}$ can in principle turn into a 
candidate fixed point with $a_{UV}>a_{IR}$ once the contribution of all free GIOs are subtracted from $a$. For this reason rather than imposing 
$a_{UV}-a_{IR}>0$ from the beginning it is better to consider every real solution stemming from maximizing the a-function.  Here we limit the analysis to the positive $R_i$ case with at least one  $R_i<1/3$ but without enforcing the bounds $b<0$ and $a_{UV}-a_{IR}<0$ or $c<0$. This is because we expect 
that if one of the $R_i<1/3$ some GIO can become free and the minimization analysis needs to be redone.
 
However, even the enlarged analysis,  didn't return potentially relevant asymptotically safe candidates. The main reason is that in all cases only few GIOs can become free making it difficult to return a positive variation of the a-function. Specifically, in most of the cases,  the smallest $R_i$ is for the 10 chiral superfield that has only 
$10\,10$ as a GIO. In few cases also other fields like $126$ or $\overline{126}$ or 
$16$ have $R<2/3$, but typically not very small, therefore it is hard to construct singlet operators with $R$-charges less than $2/3$, while in 
combination with $10$ it is hard to get many more invariants. For example the invariant $126\,10^5$ is 
antisymmetric in $10$ so with one $10$ only it vanishes.

Let's give an explicit example, i.e. the theory with the maximum among the negative 
$a_{UV}-a_{IR}$ we were able to find for all $R_i>0$. This corresponds to the theory with superpotential

\beq
\label{Wexample}
W = y_3\,10\,210\,126 + y_4\,10\,210\,\overline{126} \ .
\eeq

\noi
We find upon maximization of $a$  

\begin{align}
R &= (0.893,1.00,1.00,0.104,0.781,0.781,0.781) \ ,\\
b_{UV} &= (1.37\times10^4,27.4,7.02,7.00) \ , \\
a_{UV}-a_{IR}& =  -98.2 \ ,\quad c_{UV} = 315. \ .
\end{align}
Only the $126$ and $\overline{126}$ have $R>1$ and thus $\epsilon=-1$.  
We find that only $R_{10}=0.104$ is smaller than $1/3$ and that there is only one GIO, i.e. $10\, 10$, with the correction $\Delta(a)=a_0(2/3)-a_0(2R_{10})$ to be added to the previous $a$. 
The new maximization yields (for the solution with all $b>0$)
\begin{align}
R &= (0.897,1.00,1.00,0.103,0.777,0.777,0.777) \ , \\
b_{UV} &= (1.40\times10^4,26.8,7.15,7.13) \ ,\\
a_{UV}-a_{IR}&= -97.3 \ , \quad c_{UV} =   317. \ . 
\end{align}
Although $a_{UV}-a_{IR}$ is slightly larger it is still negative and the fixed point is excluded. Of course, also $c$ and $b$ receive small corrections. 
%	
%	Because of the 
%	
%	 Remember that any GIO contributes just as one degree of freedom to $a$ 
%	(or $c$, $b$), but large SO(10) multiplets have a huge number of d.o.f. so it is 
%	unlikely that this procedure can be successful. What would one need is 
%	that more $R$'s become close to zero, so that many GIOs become free. 
%	This does not happen in the cases studied though. 
%

So far we have investigated the case in which all $R$ were positive. However one could have one or more negative $R$-charges. This interesting case will be investigated elsewhere. 

%- - - - - - - - - - - - - - - 
%      
%- - - - - - - - - - - - - - -
%
%We return now to the solutions having one ore more negative $R_i$ charges at the candidate fixed point.  The associated operators can decouple and therefore one should consistently modify the $a$-function. For example,  let's consider  the case with the following superpotential
%\beq
%W=y_2\,210\,126\,\overline{126}+y_5\,10\,16_3^2+y_6\,\overline{126}\,16_2^2
%\eeq
%\noi
%with a candidate UV fixed point for
%\beq
%R=(2.44027,-0.23947,-0.2008,0.666667,-0.223239,1.1004,0.666667)
%\eeq
%\noi
%that passes the constraints because 
%\bea
%a_{UV}-a_{IR}&=&333.354 \ ,\quad c_{UV} = 1590.03 \ , \\ 
%b_{UV}&=&(4502.21,554.988,9.33333)   \ .
%\eea
%{\color{blue} It would good to define explicitly the U(1) charges here for the flavour anomaly. }
%The fields $126,\overline{126},16_1$ have negative $R$-charge. According to 
%\cite{Slansky:1981yr,Feger:2012bs} the products (as an example) $126^4$ include 3 gauge singlets, 
%of which one is a symmetric combination %of four single $126$'s
%\footnote{This can be checked by calculating the corresponding pleythistic exponent, 
%for a clear and useful introduction for non-mathematician see for example \cite{Hanany:2008sb} 
%and references therein.}, so this solution is invalid, and it should be corrected as proposed in \cite{Kutasov:2003iy}. 
% 
% - - - - - - - - - - -

 \subsection{On the doublet-triplet splitting problem}
Grand unified theories require the SM Higgs to arise from representations of the unified group. These contain, besides the SM Higgs weak doublet, also other states that include color triplets. In supersymmetric theories color triplet Higgses can induce dimension five supersymmetric operators mediating proton decay.  Consequently one needs to keep the color triplet very heavy, typically heavier than the GUT scale, while keeping the doublet light for the theory to be  viable. The doublet-triplet splitting (DT) problem begs the question: what keeps the doublets light and the triplets heavy? 
To ameliorate the severity of the problem several proposals have been made in the literature, such as the Dimopoulos-Wilczek mechanism that requires the introduction of an adjoint SO(10) chiral field. The field can acquire, due to its antisymmetric nature, two independent vacuum expectation values: A non-vanishing one for the color triplet and a vanishing-one for the doublet   \cite{Dimopoulos:1981xm}. This is also known as the missing VEV mechanism \cite{Dimopoulos:1981xm,Babu:1993we}. Other possibilities are the missing partner \cite{Grinstein:1982um,Masiero:1982fe,Hisano:1994fn,
Babu:2006nf,Babu:2011tw} and the 
orbifold construction \cite{Kawamura:2000ev,Altarelli:2001qj,Hall:2001pg,Hebecker:2001wq,Hebecker:2001jb,
Asaka:2001eh,Hall:2001xr,Dermisek:2001hp}. These mechanisms  require another layer of model building in SO(10) with the addition of extra fields such as the aforementioned $45$ (missing VEV), additional $126+\overline{126}$ (missing partner) or extra Kaluza-Klein states (orbifold). 

Although a more thorough analysis of the possible occurrence of an UV fixed point in models directly addressing the DT problem and low energy R-symmetry will be performed elsewhere we can already discuss a special case here.  By using the solution, found in Section \ref{W}, where all chiral fields, including the extra ones needed for the DT splitting, are constrained to have R-charge equal to $2/3$ except for the $16_1$ which is allowed to have a very large R-charge we can argue that this solution is a plausible candidate for an UV finite GUT theory.

\section{Outlook and conclusions}
\label{conclusions}
We investigated the possibility for phenomenologically motivated supersymmetric grand unified theories to feature an interacting ultraviolet fixed point before reaching the gravity transition scale. Using a set of nonperturbative tools ranging from a-maximization to the positivity of relevant  central charges we nonperturbatively rule out this possibility for a broad class of prime candidates. We have also discovered a less exotic candidate theory, passing these tests that, although features the physically relevant fields, is not yet phenomenologically viable.  Nevertheless the exotic candidates simultaneously elucidate the challenges and hint to the required underlying structure of potentially viable asymptotically safe grand unified theories. 

We focussed in this initial work on grand unified theories on the SO(10) theory but we plan to extend the analysis to similar $E_6$ realizations \cite{Bajc:2013qra,Babu:2015psa} as well as  SU(5) \cite{Grinstein:1982um,Masiero:1982fe,Hisano:1994fn}. 

\section*{Acknowledgments}
BB would like to thank CP3-Origins \& the Danish IAS for hospitality, during which this work started. FS thanks the hospitality of the Lyon high energy group  as {\it Professeur mois invite} where this work was finalized. The work of BB has been supported by the Slovenian Research Agency. FS acknowledges the support of the Danish National Research Foundation under the grant number DNRF90. 

\appendix

\end{document}